\documentclass[fleqn,usenatbib]{mnras}
\usepackage{url}
\usepackage{natbib}
\usepackage[normalem]{ulem}
\usepackage[T1]{fontenc}
\usepackage{txfonts}
\usepackage{graphicx}
\usepackage{savesym}
\savesymbol{leftroot}
\savesymbol{uproot}
\savesymbol{iint}
\savesymbol{iiint}
\savesymbol{iiiint}
\savesymbol{idotsint}
\savesymbol{Bbbk}
\usepackage{orcidlink}
\usepackage{amsmath,amssymb,physics,bm,esint}
\usepackage{xlmacros}
\usepackage[pdfpagelabels=false]{hyperref}
\usepackage[dvipsnames]{xcolor}
\usepackage{bm}
\usepackage{soul}

\DeclareRobustCommand{\VAN}[3]{#2}
\let\VANthebibliography\thebibliography
\def\thebibliography{\DeclareRobustCommand{\VAN}[3]{##3}\VANthebibliography}

\title{A Differentiable Perturbation-based Weak Lensing Shear Estimator}
\author[X. Li et al.]{
    Xiangchong Li$^{1,2}$\thanks{xiangchl@andrew.cmu.edu}\orcidlink{0000-0003-2880-5102},
    Rachel Mandelbaum$^{1}$\orcidlink{0000-0003-2271-1527},
    Mike Jarvis$^{3}$\orcidlink{0000-0002-4179-5175},
    Yin Li$^{4}$\orcidlink{0000-0002-0701-1410},
    Andy Park$^{1}$\orcidlink{0000-0002-8318-8226},
    Tianqing Zhang$^{1}$\orcidlink{0000-0002-5596-198X}\\
$^{1}$ Department of Physics, McWilliams Center for Cosmology, Carnegie Mellon
University, Pittsburgh, PA 15213, USA\\
$^{2}$ Kavli Institute for the Physics and Mathematics of the Universe (Kavli
IPMU, WPI), UTIAS, The University of Tokyo, 5-1-5 Kashiwanoha, Kashiwa,\\ Chiba
277-8583, Japan\\
$^{3}$ Department of Physics and Astronomy, University of Pennsylvania,
Philadelphia, PA 19104, USA\\
$^{4}$ Department of Mathematics and Theory, Peng Cheng Laboratory, Shenzhen,
Guangdong 518066, China
}

\date{Received July ??, 2023; accepted Month ??, 2023}
\pubyear{2023}

\begin{document}

\label{firstpage}
\pagerange{\pageref{firstpage}--\pageref{lastpage}}
\maketitle

\begin{abstract}
    Upcoming imaging surveys will use weak gravitational lensing to study the
    large-scale structure of the Universe, demanding sub-percent accuracy for
    precise cosmic shear measurements. We present a new differentiable
    implementation of our perturbation-based shear estimator (\FPFS{}), using
    \jax, which is publicly available as part of a new suite of analytic shear
    algorithms called \anacal{}. This code can analytically calibrate the shear
    response of any nonlinear observable constructed with the \FPFS{} shapelets
    and detection modes utilising auto-differentiation (\texttt{AD}),
    generalizing the formalism to include a family of shear estimators with
    corrections for detection and selection biases. Using the \texttt{AD}
    capability of \jax, it calculates the full Hessian matrix of the non-linear
    observables, which improves the previously presented second-order noise
    bias correction in the shear estimation. As an illustration of the power of
    the new \anacal{} framework, we optimize the effective galaxy number
    density in the space of the generalized shear estimators using an LSST-like
    galaxy image simulation for the ten-year LSST. For the generic shear
    estimator, the magnitude of the multiplicative bias $|m|$ is below $3\times
    10^{-3}$ (99.7\% confidence interval), and the effective galaxy number
    density is improved by 5\%\,.  We also discuss some planned future
    additions to the \anacal{} software suite to extend its applicability
    beyond the \FPFS\ measurements.
\end{abstract}


\section{INTRODUCTION}

Weak gravitational lensing refers to the  weak but coherent distortions in the
shapes of distant background galaxies caused by the foreground inhomogeneous
mass distortion. It is a unique and essential cosmological tool, providing
crucial insight into the Universe's most mysterious constituents: dark matter
and dark energy. The fundamental principle behind this tool lies in Einstein's
general theory of relativity, which posits that the presence of mass can
distort the fabric of space-time, causing light to follow curved trajectories
(see \citealt{rev_wl_Bartelmann01, rev_wlLSS_Refregier2003,
rev_wlDM_Massey2010, rev_cosmicShear_Kilbinger15} for reviews on weak lensing).

The weak lensing distortion, which is termed ``shear'', is a measure of the
anisotropic stretching of the image of a galaxy due to the intervening mass
distribution. Accurate measurement of the weak lensing shear is vital for
understanding the structure and evolution of the universe. Due to its subtle
nature, this task is inherently complex and is compounded by observational
challenges, including the point-spread function (PSF) from telescope optics and
atmospheric effects \citep{PSF_rev_Paulin2008, PSF_rev_Liaudat2023}; model bias
from unrealistic assumptions on galaxy morphology
\citep[e.g.,][]{modelBias_Bernstein10}; bias from image noise due to the
non-linearity in shear estimators \citep[e.g.,][]{noiseBiasRefregier2012}; bias
due to anisotropies in our selected galaxy sample induced by the sample
selection \citep{KSB_Kaiser1995} and detection \citep{metaDet_Sheldon2020}, and
bias arises from blending and deblending of galaxies \citep{FPFS_Li2018}. We
refer interested readers to \citet{rev_wlsys_Mandelbaum2017} for a
comprehensive review of systematics in shear estimation. Despite these
difficulties, \metadet{} \citep{metacal_Huff2017, metacal_Sheldon2017,
metaDet_Sheldon2020}, an empirical shear estimator, has been developed to
measure shear from blended galaxy images with subpercent accuracy without
relying on any calibration from external image simulations. In addition to
\metadet{}, a Bayesian method \BFD{} \citep{BFD_Bernstein2016} reaches
subpercent accuracy for isolated galaxies, but there is a residual
percent-level multiplicative bias for blended galaxies.

In addition to these shear estimation methods, there has been a long tradition
of developing purely analytic\footnote{
    By analytic here, we mean that all the formulae used to measure and
    calibrate the shear are derived directly from first principles and do not
    involve any empirical steps.
}
shear estimators \citep{Z08, Z15, FPFS_Li2018, FPFS_Li2022, FPFS_Li2023}.
\FPFS{} is a perturbation-based analytic shear estimator without any assumption
regarding galaxy morphology \citep{FPFS_Li2018}. Specifically,
\citet{FPFS_Li2018} constructs an ellipticity estimator with a set of linear
observables that is obtained by convolving the image with a set of kernels
(e.g., shapelet functions: \citealt{shapeletsI_Refregier2003,
polar_shapelets_Massey2005}) after PSF deconvolution \citep{Z08}.
\citet{FPFS_Li2018} derive the linear shear responses of the linear observables
using the shear responses of these kernels to obtain the shear response of the
ellipticity. \citet{FPFS_Li2022} corrects for second-order noise bias using the
second-order derivatives (Hessian matrix) of the ellipticity with respect to
the linear observables and the covariance matrix of the measurement error on
the linear observables induced by image noise. Finally, \citet{FPFS_Li2023}
interpreted the detection and selection processes as a weighting of image
pixels after convolution with a set of kernels, and corrected for the detection
bias in the \FPFS{} shear estimator. \citet{FPFS_Li2023} showed that the shear
estimation bias in \FPFS{} is less than $0.5\%$ of the shear signal in the
presence of blending. The \FPFS{} shear estimator is more than a hundred times
faster than the widely used \metadet{} \citep{metaDet_Sheldon2020} due to its
analytic features\footnote{
    \metadet{} applies a shear distortion to each interpolated galaxy image
    \citep{FD_sample_Bernstein2014} to derive the shear response, and the
    interpolation is computationally heavy.
}.

However, the formalism in that work  neglected a few relatively small elements
in the Hessian matrix to simplify the analytic derivation of the noise bias
correction. This leaves a residual multiplicative bias of about -0.4 percent
relative to the true shear for faint galaxies. In addition, it is too
complicated to analytically extend the ellipticity and detection / selection
weights to a general form and derive the shear response and noise bias for
them. These limitations motivate our work in this paper.

\jax{} \citep{jax_Bradbury2018} is a powerful computational framework developed
by Google that aims to offer a new perspective on numerical computing. It
extends the ability of the Python programming language to automatic
differentiation of Python and \numpy{} functions. To be more specific, we can
use a \jax{}-based framework to compute derivatives of the functions with
respect to the input variables. In this paper, we put the \FPFS{} shear
estimator into the \jax{} framework, and use the auto-differentiation
(\texttt{AD}) feature to improve the noise bias correction in the \FPFS{} shear
estimator. In addition, we generalize the \FPFS{} shear estimator to a more
generic form and improve its precision using the \texttt{AD} capabilities of
\jax{}. In order to provide a simple interface through one repository that
enables users to seamlessly update as new versions of analytic shear estimators
become available, we introduce a novel analytical framework for shear
calibration, termed
\anacal{}\footnote{https://github.com/mr-superonion/AnaCal/}. This system is
dedicated to improving weak lensing shear calibration by creating easy-to-use
tools that accurately turn measured galaxy shapes into shear distortions. For
better manageability, modularity, and scalability, we have organized our
codebase into two separate repositories. The first repository handles the
analysis of galaxy images to generate shape catalogs and is accessible at
\FPFS{} GitHub Repository\footnote{https://github.com/mr-superonion/FPFS/}. The
second repository is dedicated to calculating shear based on these shape
catalogs, and can be found at the \impt{} GitHub
Repository\footnote{https://github.com/mr-superonion/imPT/}.  However, users of
this method can simply interact with the \anacal{} repository.

In this paper, we first introduce the \FPFS{} shear estimator in
section~\ref{sec:method}. After that, we present two applications of our shear
estimator in section~\ref{sec:applications}: first, improving the noise bias
correction and second, using a more generic shear estimator to improve
precision. Finally, we summarize these results and the future outlook in
section~\ref{sec:summary}.

\section{PERTURBATION BASED SHEAR ESTIMATOR}
\label{sec:method}

\subsection{\FPFS{} shear estimator}

In this subsection, we briefly review the \FPFS{} shear estimator developed in
\citet{FPFS_Li2018, FPFS_Li2022, FPFS_Li2023}. A weak lensing shear distortion
caused by a foreground inhomogeneous mass distribution changes the light
profiles of distant background galaxies. Here
\begin{equation}
    \bm{A}=\begin{pmatrix}
    1- g_1   &  -g_2\\
    -g_2      &  1+g_1
    \end{pmatrix}
\end{equation}
is the Jacobian matrix of the mapping from the lensed sky to the true sky. The
parameters $(g_1, g_2)$ are employed to quantify shear distortion within the
framework of the celestial coordinate system. Specifically, $g_1$ accounts for
the elongation of the image in the horizontal direction, while $g_2$ is
responsible for the stretching of the image along an axis inclined at a
45-degree angle to the horizontal. In this paper, we set the lensing
convergence \citep{rev_wl_Bartelmann01} to zero to simplify the notation.

The \FPFS{} ellipticity, a spin-$2$ observable, which negates under a
90-degrees rotation (see appendix~B of \citealt{FPFS_Li2023} for more details),
is used to measure weak lensing shear.  It is typically described as a complex
number, $g = g_1 + \mathrm{i} g_2$.  For this work, it is defined with the
\FPFS{} shapelet modes (denoted as $M_{nm}$):
\begin{align}
\label{eq:ellipticity_def1}
e_1 + \mathrm{i}\, e_2 \equiv \frac{M_{22}}{M_{00}+C}\,.
\end{align}
The weighting parameter $C$, introduced by \citet{FPFS_Li2018}, is a constant
for a galaxy sample. Different choices for $C$ assign different relative weights
to galaxies with different brightnesses.
Moreover, it ensures that the noise bias correction remains finite and
manageable \citep{FPFS_Li2022}. $M_{nm}$ are polar shapelet modes measured from
the observed noisy image after PSF deconvolution in Fourier space
\citep{FPFS_Li2018}:
\begin{equation}
\label{eq:shapelet_modes}
    M_{nm} \equiv \iint \dd[2]{k} \, \grave{\chi}^{*}_{nm}(\vk)
    \frac{f^p_\vk}{p_\vk}\,,
\end{equation}
where $f^p_\vk$ is the observed (PSF-convolved, noisy) image in Fourier space
and $p_\vk$ is the PSF image in Fourier space, and $\grave{\chi}_{nm}$ is the
Fourier transform of the polar shapelet basis function with radial quantum
number ``$n$'' and angular quantum number ``$m$'' (also known as spin number).
The polar shapelet basis functions \citep{polar_shapelets_Massey2005,
Shapes_Bernstein2002} are defined as
\begin{equation}
\label{eq:shapelet_func}
\begin{split}
\chi_{nm}(\vx \,|\, \sigma_h)&=(-1)^{(n-|m|)/2}\left\lbrace
    \frac{[(n-|m|)/2]!}{[(n+|m|)/2]!}\right\rbrace^\frac{1}{2}\\
    &\times
    \left(\frac{\absolutevalue{\vx}}{\sigma_h}\right)^{|m|}
    L^{|m|}_{\frac{n-|m|}{2}}\left(\frac{\absolutevalue{\vx}^2}{\sigma_h^2}\right)
    e^{-\absolutevalue{\vx}^2/2\sigma_h^2}e^{-im\theta},
\end{split}
\end{equation}
where $L^{|m|}_{\frac{n-|m|}{2}}$ are the Laguerre polynomials and $\sigma_h$
is the smoothing scale of shapelets and the detection kernel. $n$ can be any
non-negative integer, and $m$ is an integer between $-n$ and $n$ in steps of
two. $\sigma_h$ is set to $0\farcs52$ in this paper, which optimize the
effective galaxy number density of the algorithm for simulated images
with seeing size of $0\farcs8$\,.

\begin{figure}
\centering
\includegraphics[width=0.45\textwidth]{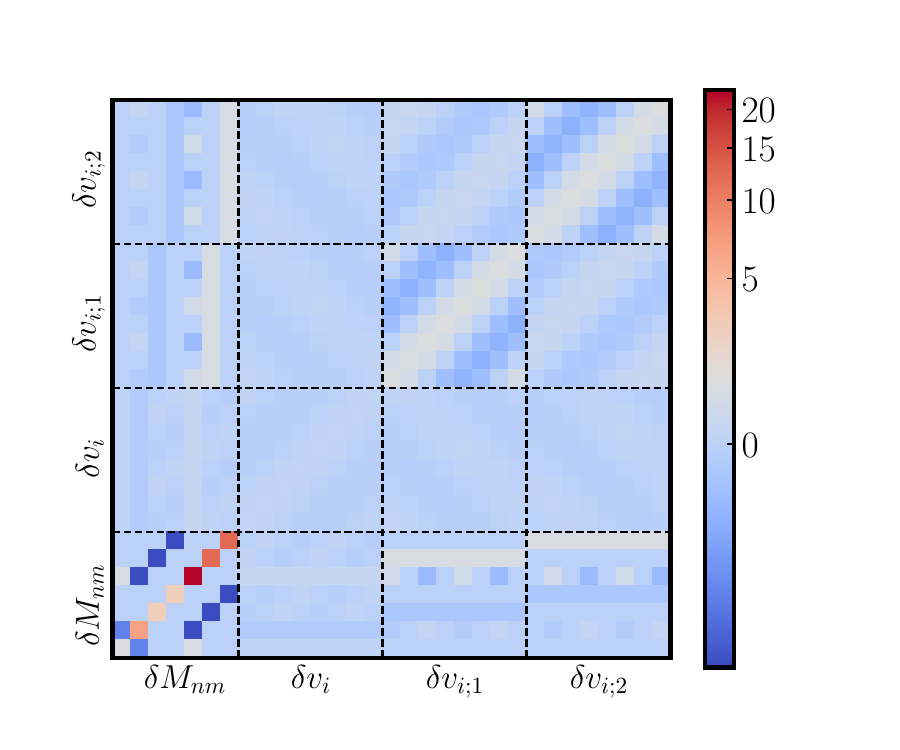}
\includegraphics[width=0.45\textwidth]{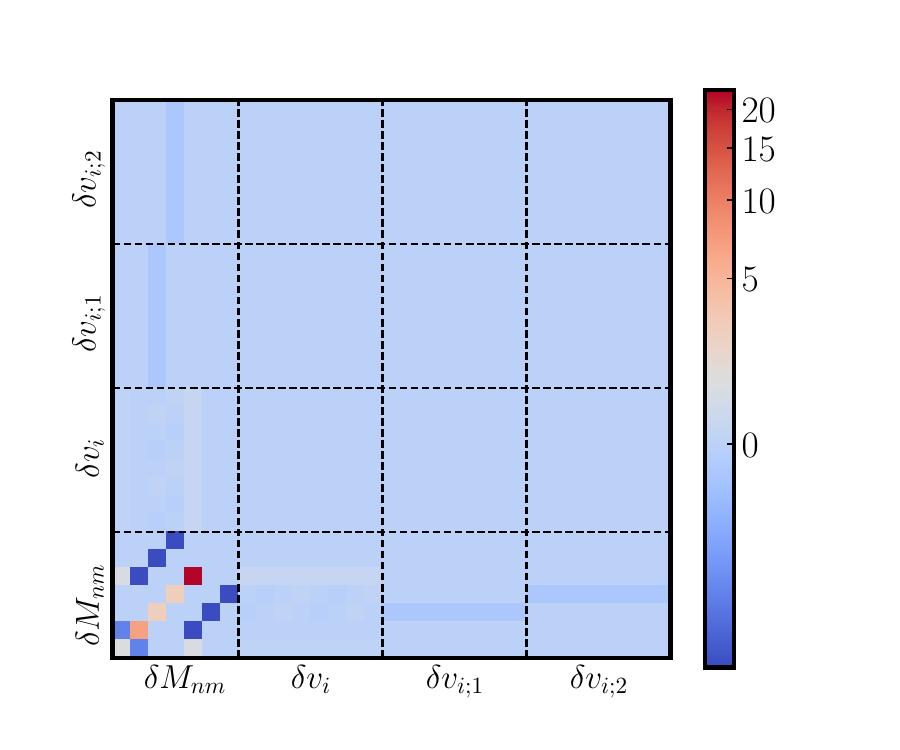}
\caption{
    The covariance matrix ($K_{v_j}^{v_i}$ in equation~\eqref{eq:ell_noisy}) of
    the measurement errors on linear observables (e.g., seven shapelet modes
    and eight detection modes with corresponding sixteen shear responses) due
    to the image noise. The shapelet modes from left to right are $M_{00}$,
    $M_{20}$, real and imaginary components of $M_{22}$, $M_{40}$, and real and
    imaginary components of $M_{42}$. As shown, the elements in the lower-left
    block have the largest magnitude dominating the noise bias correction. The
    upper panel shows the full covariance matrix, and the lower panel shows the
    covariance matrix used for the analytic noise bias correction in
    \citet{FPFS_Li2023}, which has several sub-blocks that were set to 0.
    }
    \label{fig:covmat}
\end{figure}

In addition to shapelets, we use eight detection modes ($v_i$, where $i=0\dots
7$) to detect galaxies, and the detection modes are computed for each pixel as
\begin{equation}
\label{eq:det_modes}
    v_i = \iint \dd[2]{k}\, \psi^*_{i}(\vk) \frac{f^p_\vk}{p_\vk}\,,
\end{equation}
where the coordinate center is set to the pixel center. The detection kernels
$\psi^*_{i}(\vk)$ are defined in Fourier space:
\begin{equation}
\label{eq:det_kernel}
\psi^*_{i}(\vk) = \frac{1}{(2\pi)^2}e^{-\abs{\vk}^2\sigma_h^2/2}
\left( 1 - e^{ \rmi \vk \cdot \vx_i} \right)\,,
\end{equation}
where $\vx_i$ are vectors with length equal to the image pixel side length and
orientations pointing towards eight directions ($i=0\dots 7$) separated by $\pi
/ 4$\,. The detection modes measure the difference in brightness between this
pixel and its eight adjacent pixels. In practice, we convolve images with the
detection kernels to compute detection modes for each pixel, and use a smooth
step function to select peaks with all of these detection modes above a
threshold as galaxy candidates. In our current detection algorithm, galaxies
with centers proximate to pixel edges or corners may be identified twice or
four times, respectively. However, each of these detections receives a lesser
weight, approximately  0.5 or 0.25 depending on the number of detections.
Furthermore, the centroid of each detection is located at the peak pixel's
center. Given that the shear response of centroid refinement cannot be
analytically derived with our current methods, it prevents us from refining the
estimated galaxy centroid. We intend to delve deeper into these aspects and
optimize them in future work on the detection algorithm.

More specifically, we use a galaxy weight $w$ to select galaxies with flux and
size above corresponding thresholds \citep{FPFS_Li2022}, and detect galaxies
from images \citep{FPFS_Li2023}. The galaxy weight is required to be
second-order differentiable for the noise bias correction \citep{FPFS_Li2023}.
The weight can be constructed as a product of either truncated sine functions
(see equation~(46) of \citealt{FPFS_Li2023}) or truncated polynomial functions
\citep{smoothstep_Hazimeh20} of shapelet modes $M_{nm}$ and eight detection
modes $v_i$ ($i = 0, \cdots, 7$):
\begin{equation}
    w = T^\mathrm{sel}_0\!\left(M_{00} \right)\,
    T^\mathrm{sel}_2\!\left(M_{00} + M_{20} \right)\,
    \prod_{i=0}^{7} T^\mathrm{det}(v_i)\,,
\end{equation}
where $T^\mathrm{sel}_0$ is used to select bright galaxies with high
signal-to-noise ratio (SNR), $T^\mathrm{sel}_2$ is used to select well-resolved
large galaxies, and $T^\mathrm{det}$ is used to detect galaxies. Effectively,
we assign a weight to each pixel, and the weight reflects the possibility of
the peak being the center of a galaxy.

Our detection algorithm is a simplified version of the one used by
\texttt{SExtractor} \citep{Sextractor_Bertin1996}. It operates by identifying
peak pixels that exceed the sky background threshold. However, unlike the
\texttt{SExtractor} peak detection algorithm, our method does not employ an
adaptive process to refine galaxy centers and galaxy boundaries. This is an
``ambiguous detection'', where detected peaks can have larger offsets from true
galaxy center compared to detections with centroid refinement. Despite this
limitation, our algorithm excels in providing an accurate analytic response to
shear. This response is crucial for correcting biases in shear estimation,
making our approach both effective and reliable in this respect.

\begin{figure*}
\centering
\includegraphics[width=0.90\textwidth]{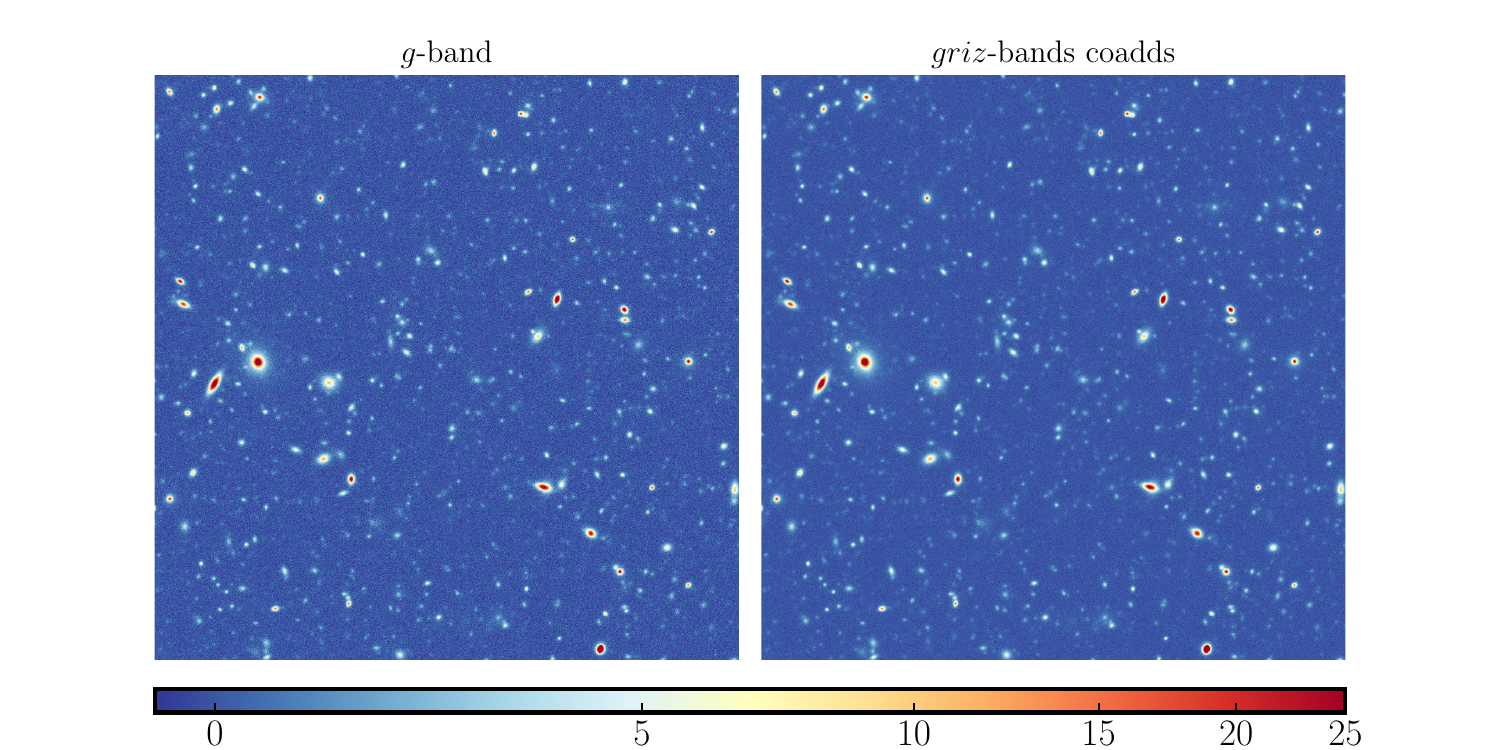}
\caption{
    The $g$-band image (left), which has the highest SNR, and the coadded image
    of $griz$-bands (right) of the LSST-like image simulation
    \citep{metaDet_LSST2023} of the same region using the expected ten-year
    LSST noise level. The region is a random cut-out from the simulated
    exposure, and it covers $3.4 \times 3.4$ square arcmins. The images are
    coadded with inverse variance weight based on background photon noise.
}
\label{fig:simulation}
\end{figure*}

As shown in equation~(9) of \citet{FPFS_Li2023}, assuming that the histogram of
image noise is symmetric with respect to zero after subtracting the background
level, the expectation value of the noisy sheared observable (with a tilde) is
related to the noiseless sheared one (without accent tag) as
\begin{equation}
\label{eq:ell_noisy}
\begin{split}
\langle \tilde{w} \tilde{e}_\mu\rangle= \langle w e_\mu \rangle
+ \frac{1}{2}\sum_{i,j} \left\langle \frac{\partial^2
    (\tilde{w} \tilde{e}_\mu)}{\partial q_i \partial q_j}
    K_{q_j}^{q_i} \right\rangle
+ \mathcal{O}\left((\delta q_i)^4\right)\,,
\end{split}
\end{equation}
and, following equation~(6) of \citet{FPFS_Li2023}, the expectation value of
the sheared noiseless observable (without accent tag) is related to the
intrinsic unsheared noiseless one (with a bar) as
\begin{equation}
\label{eq:ell_sheared}
\begin{split}
\langle w e_\mu\rangle= \langle \bar{w} \bar{e}_\mu \rangle
+ \sum_{\rho=1,2} \sum_{i}
g_\rho \left\langle \frac{\partial w e_\mu }{\partial q_i}
q_{i; \rho} \right\rangle
+ \mathcal{O}\left(g^3\right)\,.
\end{split}
\end{equation}
$q_{i; \rho} \equiv \partial q_i / \partial g_\rho$\, is the shear response of
the linear observable $q_i$\,. $K_{q_j}^{q_i} \equiv \langle \delta q_i \delta
q_j\rangle_\text{n}$ is the covariance matrix of the linear observables that
are used by the \FPFS{} shear estimator, which is demonstrated in
Fig.~\ref{fig:covmat}. Here $\langle \bigcdot \rangle$ is the average over
noiseless or noisy galaxies; whereas $\langle \bigcdot \rangle_\text{n}$ is the
average over noise realizations. In Equations~\eqref{eq:ell_noisy} and
\eqref{eq:ell_sheared}, $q_i$ are the array of shapelets and detection modes.
We note that, in principle, $q_i$ can be any linear observables as long as we
derive the shear responses and noise covariance matrix for these linear
observables. We refer interested readers to \citet{FPFS_Li2023} for a more
detailed discussion.

It is worth noting that, when deriving the algorithm, we operate under the
assumption that PSFs are oversampled. This implies that galaxies, as extended
objects, are also  oversampled. According to the Nyquist-Shannon
sampling theorem, a continuous signal can be accurately reconstructed from its
sampled, discrete counterpart without any information loss, and the signal
amplitude beyond the Nyquist wave number of our sampling remains sufficiently
low in Fourier space. Consequently, after sampling the continuous signal, there
is no aliasing bias on the Discrete Fourier Transform of the sampled signal
below the Nyquist wave number. This assumption is generally valid and practical
for ground-based astronomical observations. In practice, we compute the DFT
directly from images using a sufficiently large postage stamp size so that the
wave-number resolution is adequately high to ensure that the integrals in
equations~\eqref{eq:shapelet_modes} and \eqref{eq:det_modes} can be effectively
approximated by a Riemann sum. Furthermore, it's important to note that image
noise, which arises due to the quantum nature of light, can actually be
undersampled even in ground-based images. This is because image noise is not
smeared by the PSF, leading to its susceptibility to the aliasing effect in
observed images. For accurate noise bias correction, it becomes essential to
estimate the two-point correlation function of the aliased noise. This
estimation is derived from blank pixels that do not contain any detections.
Under the condition that PSFs are undersampled for single exposure images of
space missions, both the observed stellar images and galaxy images experience
aliasing bias. This bias predominantly affects large $|k|$ near the Nyquist wave
number in Fourier space. It is important to note that this aliasing effect
cannot be straightforwardly eliminated through PSF deconvolution. We refer
readers to \citet{metacal_Euclid2021} and \citet{metacal_Roman2023} for testing
shear estimation code on undersampled image simulations that are representative
to single exposure images from space missions. However, we are planning to
apply the algorithm to coadded images derived from single exposure images, and
we refer readers to \citet{Hirata2023} and \citet{Yamamoto2023} for deriving
oversampled coadded images from single exposure images.

However, since the formalism in \citet{FPFS_Li2023} work neglected a few
relatively small elements in the Hessian matrix to simplify the analytic
derivation of noise bias correction, the final results showed a residual bias
of about -0.4 percent relative to the true shear for faint galaxies. Here we
use $\jax{}$'s \texttt{AD} feature to automatically derive the full noise bias
correction, to extend the ellipticity and detection / selection weights to a
general form, and to derive the shear response and noise bias for them.

\subsection{Shear Response and Noise Bias Functionals}

Following \citet{FPFS_Li2023}, we correct for noise bias in the weighted
ellipticity using equation~\eqref{eq:ell_noisy}, and the noise bias functional
is denoted as $S$, where
\begin{equation}
    S\!\left(w e_\mu \right) = \frac{1}{2}\sum_{i,j}
    \frac{\partial^2 (w e_\mu)}
        {\partial v_i \partial v_j} K_{v_j}^{v_i}\,.
\end{equation}
The noise bias functional returns the noise bias correction function for an
input function of weighted ellipticity. Additionally, the functionals for the
two diagonal elements of the shear response matrix of the weighted ellipticity are
given by
\begin{equation}
R_{\mu}\!\left(w e_\mu \right) = \sum_{i}
\frac{\partial w e_\mu}{\partial v_i} v_{i; \mu} \,,
\end{equation}
where $\mu = 1, 2$\, \citep{FPFS_Li2023}. The two off-diagonal terms in the
response are spin-4 and therefore have zero mean. Moreover, following
\citet{FPFS_Li2023}, we can use equation~\eqref{eq:ell_noisy} to correct for
noise bias in the shear response measurement, and the correction term is
$S(R_\mu)$\,. Finally, the shear estimator is given by
\begin{equation}
    \label{eq:shear_est_1}
    \hat{g}_\mu = \frac{\left \langle \tilde{w} \tilde{e}_\mu
    - S\!\left(\tilde{w} \tilde{e}_\mu\right)\right\rangle}
    {\left \langle R_\mu\!\left(\tilde{w} \tilde{e}_\mu \right) -
    S\!\left(R_\mu\!\left( \tilde{w} \tilde{e}_\mu\right)\right) \right\rangle}\,.
\end{equation}
The numerator and the denominator in equation~\eqref{eq:shear_est_1} are
evaluated and averaged over the galaxies in the galaxy sample. Similar to
\citet{FPFS_Li2022, FPFS_Li2023}, we use $\widehat{w e_\mu} = \tilde{w}
\tilde{e}_\mu - S\!\left(\tilde{w} \tilde{e}_\mu\right)$ and $\widehat{R_\mu} =
R_\mu\left(\tilde{w} \tilde{e}_\mu \right) - S\!\left(R_\mu\left( \tilde{w}
\tilde{e}_\mu\right)\right)$ to denote the corresponding observables after
noise bias correction.

In this paper, we put the shear response functional ($R$) and the noise bias
functional ($S$) into the \jax{} ecosystem, in order to utilise the \texttt{AD}
in \jax{} to derive the shear response and noise bias correction functions for
any definition of ellipticity, selection and detection.

\section{APPLICATIONS}
\label{sec:applications}

In this section, we present two applications of the \texttt{AD} code to the
\FPFS{} method, resulting in an improved shear estimator within the \anacal{}
framework. In section~\ref{subsec:ap_noise}, we improve the noise bias
correction by including all the terms in the second-order noise bias
correction. In section~\ref{subsec:ap_ext}, we extend the \FPFS{} ellipticity
to a more generic form and find the optimal shear estimator from among the more
general family of estimators considered.

We quantify the bias in the shear estimation using multiplicative ($m_{1,2}$)
and additive ($c_{1,2}$) biases \citep{shearSys_Huterer2006}. Specifically, the
estimated shear, $\widehat{g}_{1,2}$, is related to the true shear, $g_{1,2}$,
as
\begin{equation} \label{eq:shear_biases}
    \widehat{g}_{1,2}=(1+m_{1,2})\,g_{1,2}+c_{1,2}\,.
\end{equation}

The LSST-like image simulation used in this paper is introduced in
\citet{metaDet_LSST2023}, and we give a brief review in appendix~\ref{app:sim}.
We show a simulated $g$-band and a $griz$-band coadded image with the ten-year
LSST noise level in Fig.~\ref{fig:simulation}. In this work, we use the coadded
LSST ten-year images in $griz$ bands to perform detection and shear
estimation\footnote{We note that LSST has six bands ($ugrizy$). However, due to
lesser SNR and challenges in modeling the PSF, we do not use the LSST $u$- or
$y$-band images.}.
Our image simulations are divided into $5000$ subfields corresponding to $590$
square degrees in total. We use all of them for section~\ref{subsec:ap_noise}
and a subset 100 of them for section~\ref{subsec:ap_ext}.

Although the simulation package \citep{metaDet_LSST2023} can be used to
generate images with stars, bad pixels, bright-star masks and image coaddition,
we focus on testing our updated shear estimation algorithm with the basic
galaxy-only image simulations in this paper. We will test our algorithm with
more realistic setups (including these additional systematics) in future work.

In the absence of explicit specification, this paper adheres to the optimal
setups for the estimators in equations~\eqref{eq:ellipticity_def1} and
\eqref{eq:ellipticity_def2}, where $\sigma_h = 0\farcs52$\,, $C=7.6$ for
original \FPFS{} shear estimator and $c_0=6$, $c_2=25$, $\alpha=0.83$,
$\beta=0.18$ for the generic \FPFS{} shear estimator introduced in
section~\ref{subsec:ap_ext}.

\subsection{Second Order Noise Bias Correction}
\label{subsec:ap_noise}

\begin{figure}
\begin{center}
    \includegraphics[width=0.48\textwidth]{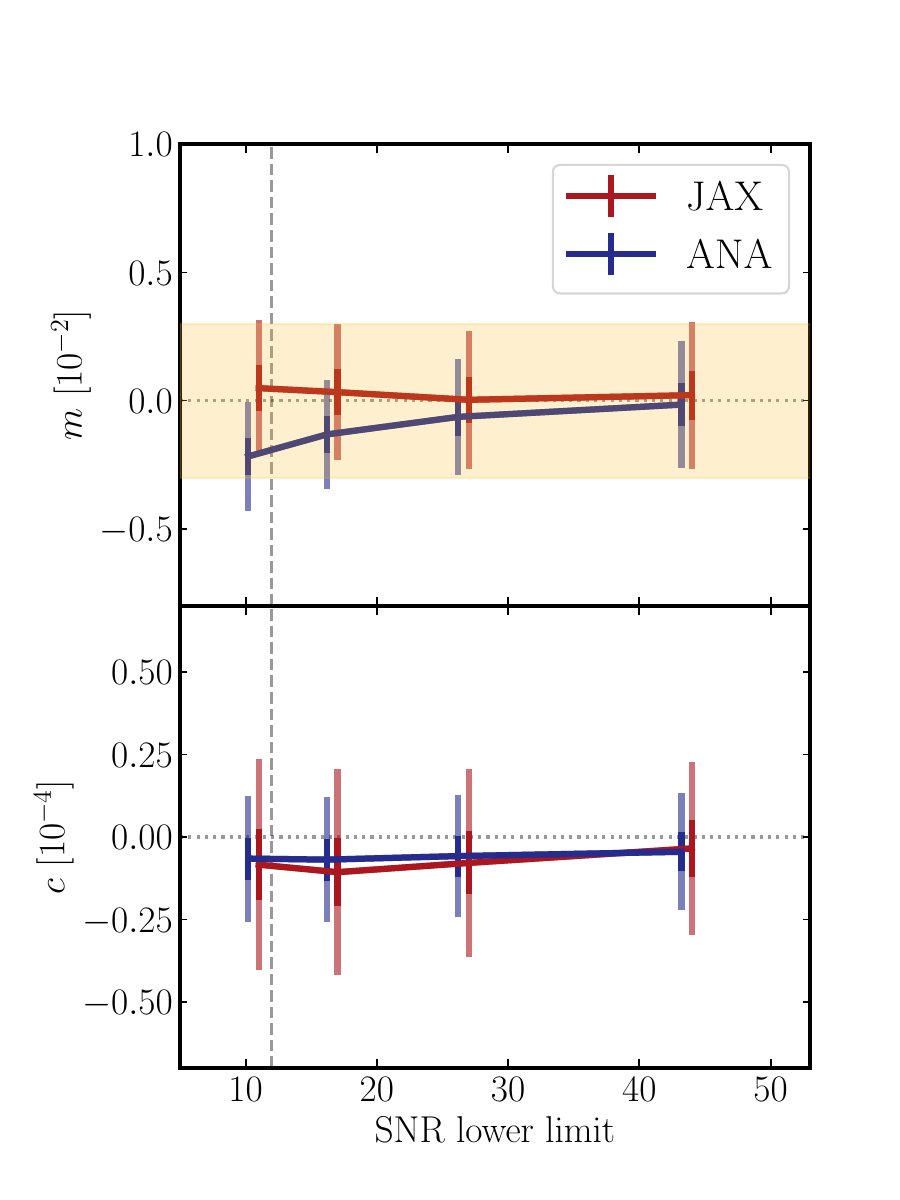}
\end{center}
\caption{
    The multiplicative bias (upper panel) and additive bias (lower panel) of
    the \FPFS{} shear estimator defined in equation~\eqref{eq:ellipticity_def1}
    with $C = 7.6$\,. Red lines are the full noise bias correction with \jax{},
    and blue lines are the analytic noise bias correction (labeled as ANA). The
    errorbars show the $1\sigma$ and $3\sigma$ uncertainties, and the shaded
    region is the LSST ten-year requirement on  control of residual
    multiplicative biases in shear \citep{LSSTRequirement2018}. Use of the full
    noise bias correction from \jax{} makes the shear estimator noisier, but
    its multiplicative bias is consistently below the LSST requirement (99.7\%
    confidence level). In this paper, we use SNR$>12$ as our default selection,
    as is indicated with the vertical dashed line.
    }
    \label{fig:fpfs_jva_mc}
\end{figure}

In \citet{FPFS_Li2023}, we analytically derived the shear responses $R_\mu(w
e)$, the noise bias corrections $S(w e)$, and $S\left(R_\mu(w e)\right)$ for
the \FPFS{} weighted ellipticity. However, when deriving the noise bias
correction, we set many elements in the covariance matrix of measurement error
(as shown in the bottom panel of Fig.~\ref{fig:covmat}) to zero to simplify our
derivation. We found a multiplicative bias in the estimated shear ranging from
$-0.3\%$ to $-0.5\%$ with 1-$\sigma$ error of $\sim$$0.1\%$, and the magnitude
of the bias increases as the image noise level increases, which suggests that
the bias originates from insufficient noise bias correction.

In this section, we compare the multiplicative biases between the analytic
version of our \FPFS{} shear estimator and the \jax{} version on the LSST-like
image simulations. We note that the difference between these two codes is that
the \jax{} version (publicly available within the \anacal{} framework) includes
the complete second-order noise bias correction while the analytic version
neglects several terms in the covariance matrix of the measurement error.

Fig.~\ref{fig:fpfs_jva_mc} shows the multiplicative  and additive biases as
functions of the SNR cut at the lower end. The resolution cut is set to $0.05$
and the lower end following \citet{FPFS_Li2023}. We find that there is a
negative multiplicative bias for the analytic correction, which is consistent
with the findings in \citet{FPFS_Li2023}. In contrast, for the \jax{}
correction, the multiplicative bias is consistent with zero, and its magnitude
is less than $0.3\times 10^{-3}$, the LSST ten-year requirement.

\subsection{Extension of \FPFS{}}
\label{subsec:ap_ext}
\begin{figure}
\begin{center}
    \includegraphics[width=0.48\textwidth]{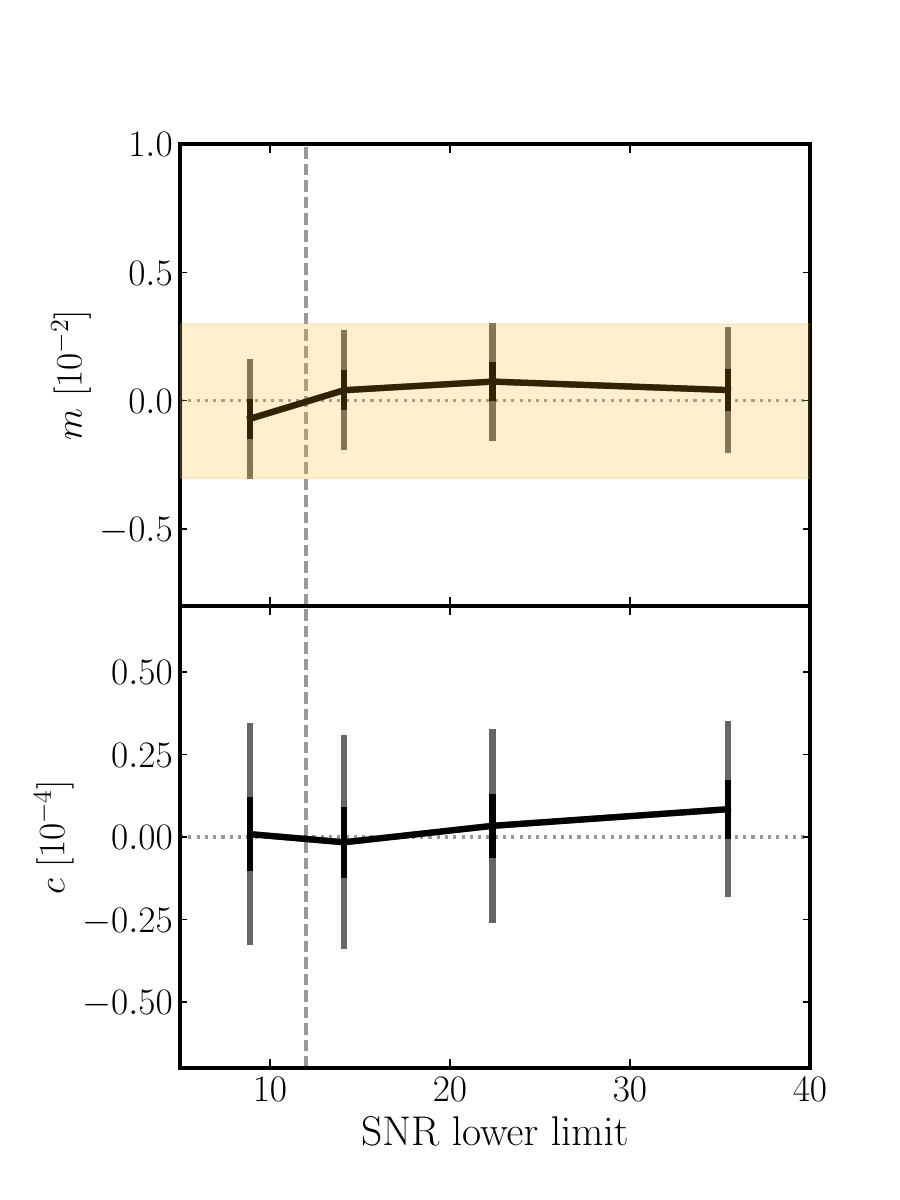}
\end{center}
\caption{
    The multiplicative bias (upper panel) and additive bias (lower panel) of
    the generalized \FPFS{} shear estimator defined in
    equation~\eqref{eq:ellipticity_def2} with the optimal parameters: ($c_0=6$,
    $c_2=25$, $\alpha=0.83$, $\beta=0.18$). The errorbars show the $1\sigma$
    and $3\sigma$ uncertainties, and the shaded region is the LSST ten-year
    requirement on control of residual multiplicative biases in shear. The
    multiplicative bias for the generic \FPFS{} shear estimator is consistently
    below the LSST requirement (99.7\% confidence level).
    }
    \label{fig:fpfs_ext_mc}
\end{figure}

To improve the precision of shear estimation with an optimal weighting scheme,
we extend the \FPFS{} ellipticity in equation~\eqref{eq:ellipticity_def1} to a
more generic form:
\begin{equation}
    \label{eq:ellipticity_def2}
    e_1 + \mathrm{i}\, e_2 \equiv
    \frac{M_{22}}{(M_{00} + c_0 \sigma_{0})^\alpha
    (M_{00} + M_{20} + c_2 \sigma_{2})^\beta},
\end{equation}
where $\sigma_{0}$ and $\sigma_{2}$ are the $1\sigma$ measurement errors on
$M_{00}$ and $M_{00} + M_{20}$ due to image noise. $\alpha$ and $\beta$ are
power-law exponents. If we set $c_0 = C / \sigma_{0}$, $c_2 = 25$, $\alpha = 1$
and $\beta = 0$ \footnote{Note, once $\beta=0$, $c_2$ does not matter in the
ellipticity definition.}, the extended ellipticity defined in
equation~\eqref{eq:ellipticity_def2} reduces to the original \FPFS{}
ellipticity defined in equation~\eqref{eq:ellipticity_def1}. We note that this
extension of the \FPFS{} ellipticity gives us more freedom to adjust the
relative weights between galaxies with different properties. We can use the
\jax{} framework to  minimize the statistical error (including intrinsic shape
noise and shape measurement error) in the extended hyperparameter space
($\alpha$, $\beta$, $c_0$, $c_2$).

\begin{figure}
\begin{center}
    \includegraphics[width=0.48\textwidth]{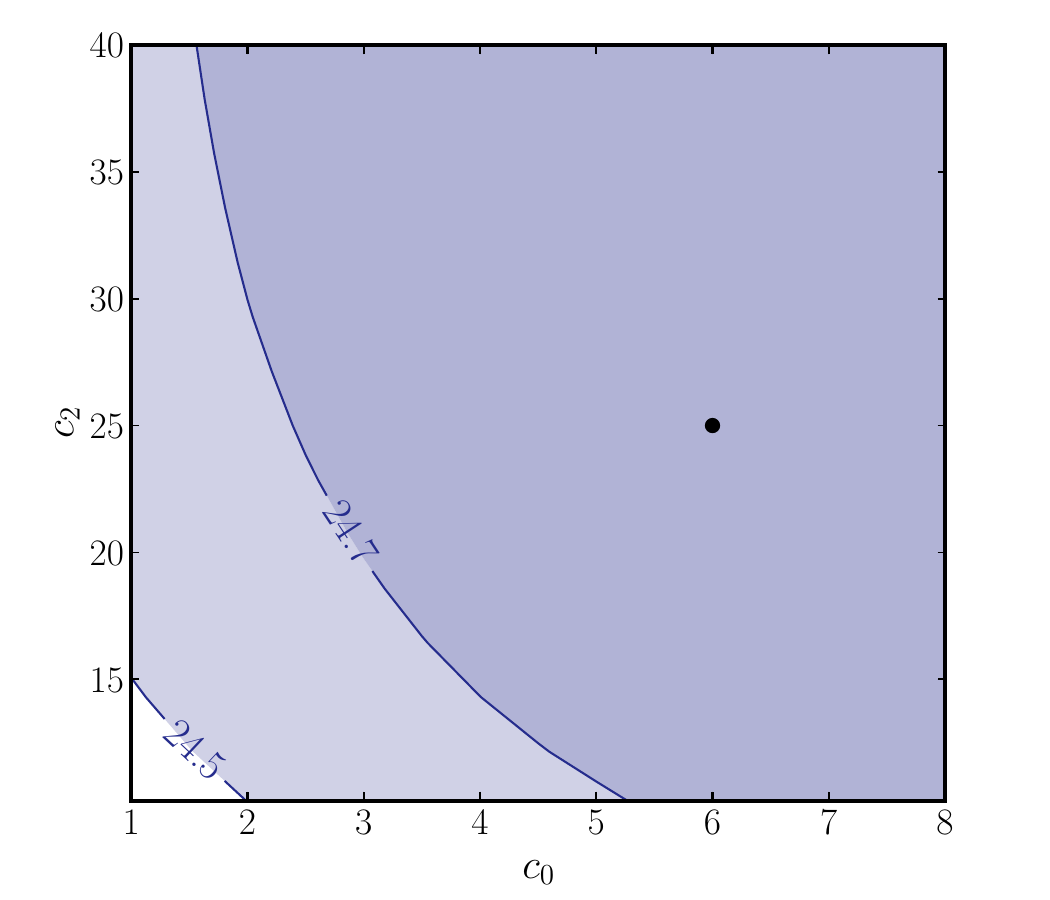}
\end{center}
\caption{
    The effective galaxy number density as a function of $c_0$ and $c_2$ for a
    fixed SNR lower limit of 12. We refer readers to
    equation~\eqref{eq:ellipticity_def2} for the definitions of $c_0$ and
    $c_2$\,. The optimal configuration is indicated with a black point. We fix
    the other two hyperparameters to the optimal ones: $\alpha=0.83$,
    $\beta=0.18$\,.
    }
    \label{fig:neff_c0c2}
\end{figure}

\begin{figure}
\begin{center}
    \includegraphics[width=0.48\textwidth]{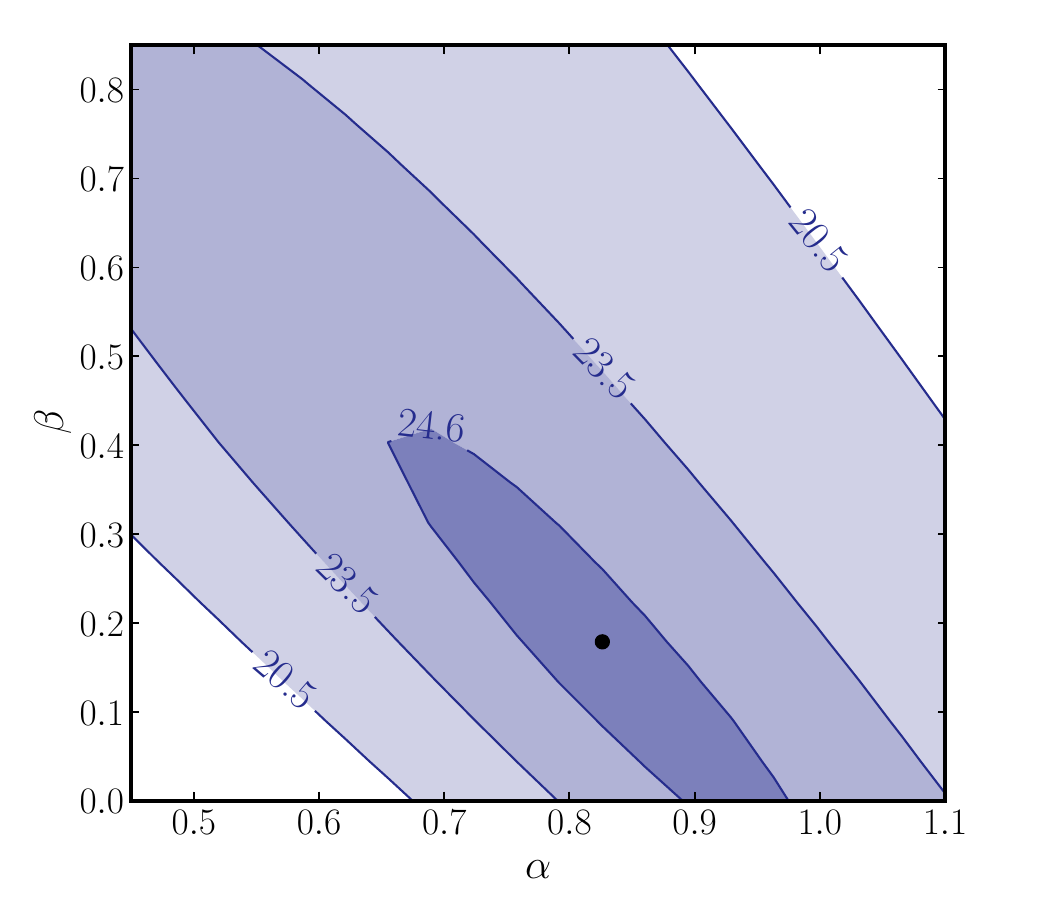}
\end{center}
\caption{
    The effective galaxy number density as a function of $\alpha$ and
    $\beta$\,. We refer readers to equation~\eqref{eq:ellipticity_def2} for the
    definitions of $\alpha$ and $\beta\,$. The optimal configuration is
    indicated with a black point. We fix the other two hyperparameters to the
    optimal ones: $c_0=6$, $c_2=25$\,.
    }
    \label{fig:neff_ab}
\end{figure}

To study the dependency of the statistical error on the hyperparameters and
find the minima of the statistical error function, we use $100$
subfields of simulated noisy blended galaxy images (see
Fig.~\ref{fig:simulation} for an example). Each simulated image has different
realizations of galaxy positions, galaxy population and image noise
realizations, and it covers 0.12 square degrees. We measure shear from each
image following the process in Appendix~\ref{app:sim} and use these $100$ shear
measurements to derive the statistical error. After normalizing the statistical
error according to the area of each simulated image, we obtain the per
component statistical error on shear estimation for a region of one square
$\mathrm{arcmin}$, which is denoted as $\sigma_g$. Then the effective number
density is estimated as
\begin{equation}
    \label{eq:effective_number}
    n_\text{eff} = \left(\frac{0.26}{ \sigma_g} \right)^2~[\mathrm{arcmin}^{-2}],
\end{equation}
where $0.26$ is the per component root-mean-square (RMS) of intrinsic shape
noise for the \reGauss{} shear estimation method
\citep{Regaussianization_Hirata2003} that is widely used in the weak lensing
community. Although different methods have different shape noise RMS, we
normalize the effective number density with the \reGauss{} intrinsic shape
noise RMS (0.26) so that we can compare our estimation of effective number
density with other methods that adopt the same normalization and the
predictions using \reGauss{} \citep[e.g.,][]{LSSTRequirement2018}. In summary,
we first estimate the statistical uncertainty in shear estimation for a one
square arcmin region. Then we derive the equivalent galaxy number for the shear
if it were to be estimated with \reGauss{}. Note that we do not estimate the
effective number density by counting the galaxy number (with weights) after
detecting galaxies from images \citep{WLsurvey_neffective_Chang2013,
LSSTRequirement2018}; that process assumes that the statistical error in shear
estimation for each galaxy is uncorrelated. Finally, we use the Nelder-Mead
minimizer \citep{NeldMead65} implemented in \texttt{scipy}
\citep{scipy_Virtanen2020} to find the minimum of the statistical error
($\sigma_g$), or equivalently the maximum of the effective number density
($n_\text{eff}$) --- which occurs at parameter values $c_0=6$, $c_2=25$,
$\alpha=0.83$, $\beta=0.18$\,.  At this point, the effective galaxy number
density is $24.8~\mathrm{arcmin}^{-2}$. During the optimization, we fix the SNR
cut to $12$, and the resolution cut to $0.05$ to select galaxies that are
sufficiently bright and  well-resolved for shear inference.

\begin{figure}
\begin{center}
    \includegraphics[width=0.48\textwidth]{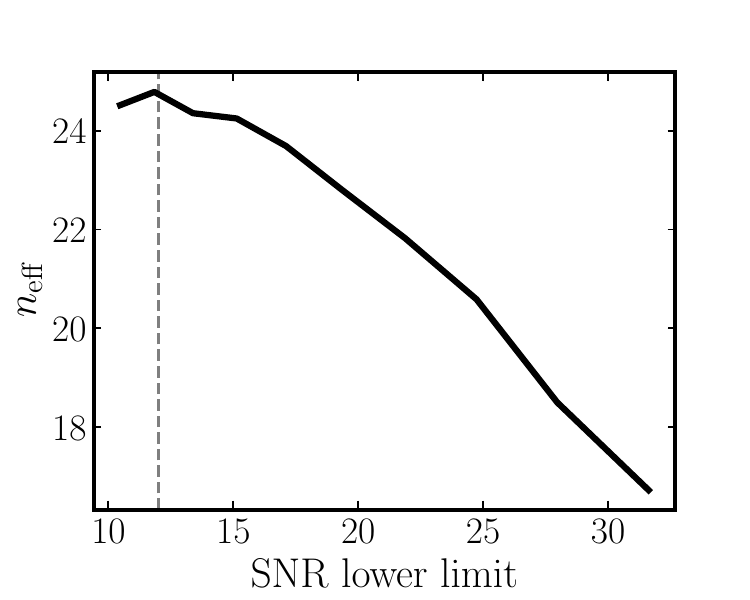}
\end{center}
\caption{
    The effective galaxy number density as a function of SNR cut. We set the
    hyperparameters to $c_0=6$, $c_2=25$, $\alpha=0.83$, $\beta=0.18$\,,
    corresponding to those for the optimal shear estimator identified in this
    work. The vertical dashed line indicates the default SNR cut used in this
    paper.
    }
    \label{fig:neff_snr}
\end{figure}

In Fig.~\ref{fig:fpfs_ext_mc}, we show the multiplicative and additive biases
for the estimator with the optimal hyperparameters. We find that the magnitude
of the multiplicative bias is consistently below $0.3\%$, the LSST science
requirement \citep{LSSTRequirement2018}. The magnitude of the additive bias is
consistently below $5\times 10^{-5}\,$.

In Figs.~\ref{fig:neff_c0c2} and~\ref{fig:neff_ab}, we show the $n_\text{eff}$
on $2$D slices through the high-dimensional hyperparameter space, corresponding
to the ($c_0$, $c_2$) and ($\alpha$, $\beta$) planes. The maximum of the
effective number density is $\sim$$5\%$ larger than that for the original
\FPFS{} shear estimator. We find that the effective number density is much more
sensitive to the choice of power-law indices ($\alpha$, $\beta$) than to the
constants ($c_0$, $c_2$) since the power law indices have a more significant
influence on the relative weights between galaxies with different brightnesses.
Additionally, the effective number density varies rapidly in the
$\beta=2\,\alpha$ direction in the ($\alpha$, $\beta$) plane, and it varies
slowly in the orthogonal direction. Moreover, the \jax{} framework facilitates
the flexibility of the shear estimator by allowing both the weight and
ellipticity to be defined as any functions of the linear modes. This capability
expedites the exploration and optimization of effective number density within
the space of possible functions while still ensuring the method is unbiased.

Additionally, we show the effective number density as a function
of SNR cut in Fig.~\ref{fig:neff_snr}. We find that for an SNR lower limit
exceeding $15$, the effective number density significantly decreases. However,
the plot shows that there is relatively little value in decreasing the SNR
lower limit well below 15, since galaxies at this faint end are dominated by
the measurement error caused by image noise.  This finding explains our choice
of a default lower limit of SNR$>12$. However, we note that the absolute value
of the effective number density may not be realistic since we have not
precisely calibrated the number density in the simulations to any real dataset.
We primarily focus, therefore, on relative changes in the number density.
Figs.~\ref{fig:neff_c0c2}--\ref{fig:neff_snr} are used to demonstrate the
possibility of optimizing the effective number density from the image level,
but not to predict the effective number density of the LSST final year survey.

To make sure that our shear estimator is stable, we show the multiplicative
bias on the $1$D slice lines through this four-dimensional parameter space in
appendix~\ref{app:stable}. We find that the multiplicative bias is not
significantly dependent on the hyperparameters, and it is less than
$3\times10^{-3}$ within a wide range of values in the hyperparameter space.

\subsection{Benchmarking}
We benchmark the code on a Intel(R) Xeon(R) Gold 6240R CPU @ 2.40GHz using a
simulated region covering one square degree with ten-year LSST depth to
compress images to catalogs, and report the results as follows:
\begin{enumerate}
    \item The \FPFS{} code (\url{https://github.com/mr-superonion/FPFS/}) takes
        $0.18$ CPU hour to detect and process the images and generate shear
        catalogs;
    \item The \impt{} code (\url{https://github.com/mr-superonion/imPT/})
        takes $0.045$ CPU hour to derive shear from the catalogs.
    \item The analytic code in \citet{FPFS_Li2023} takes $0.004$ CPU hours to
        derive shear from the catalogs.
\end{enumerate}
In summary, with the \jax{}-based version of \FPFS{} that is available through
\anacal{}, it would take less than four thousand CPU hours to infer shear from
the LSST ten-year coadds across the entire survey area.

\section{SUMMARY AND OUTLOOK}
\label{sec:summary}

In this paper, we show the results of implementing the \FPFS{} shear estimator
within the \jax{} ecosystem to utilise the automatic differentiation
(\texttt{AD}) functionality in \jax{} to improve the second-order noise bias
correction. Furthermore, we extend the \FPFS{} shear estimator to a generic
form and use the \texttt{AD} to derive shear response and noise bias correction
for that estimator.

We test the algorithm with the LSST-like image simulations
\citep{metaDet_LSST2023} using the coadded image in the $griz$-bands, and we
report that the effective number density is improved by $\sim$$5\%$ compared to
the original \FPFS{} shear estimator. We find that the magnitude of the
multiplicative bias is consistently less than $0.3\%$\, (LSST ten-year
requirement on the control of multiplicative bias) within the $3\sigma$
uncertainties, illustrating the promise of this shear estimator for LSST.

In the future, as a further step towards a practical use of the application in
real survey data, we will apply the shear estimator to image simulations with
stars, bad pixels, and bright star masks as in \citet{metaDet_LSST2023}. This
will enable confirmation that its performance carries over even in the presence
of these additional challenges, along with opportunities to further develop the
method to face any issues that this exercise reveals.

In order to provide a simple interface through one repository that enables
users to seamlessly update as new versions (even with substantially different
formalism) become available, we introduce a novel analytical framework for
shear calibration, termed \anacal{}. This framework is devised to bridge
various analytic shear estimators that have been developed (e.g.,
\texttt{FPFS}) or are anticipated to be created in the future. We intend to
develop a suite of analytical shear estimators capable of inferring shear with
subpercent accuracy, all while maintaining minimal computational time. Below,
we outline several possible avenues for future shear estimation method
development.
\begin{itemize}
    \item Enhancing Shear Estimation through Higher-Order Moments: Typically,
        the shear estimation process employs the lowest-order (specifically,
        second-order) spin-2 moments. Nonetheless, as suggested in
        \citet{shapeletsII_Refregier2003}, the utilization of fourth-order
        moments holds potential for advancing shear inference. We plan to
        incorporate this fourth-order shear estimator into the \FPFS{}
        framework, aiming to amplify the SNR in shear inference by leveraging
        both second and fourth-order moments while providing additional
        opportunities for PSF systematics mitigation.
    \item Adaptive Kernel Shear Estimator: The existing \FPFS{} shear estimator
        utilizes a fixed-kernel approach; more specifically, the kernel
        (namely, the Gaussian kernel in shapelets) does not dynamically adjust
        according to the size and shape of each observed galaxy. In future
        work, we aim to augment the precision of shear estimations by
        developing an adaptive kernel approach (e.g.,
        \citealt{Regaussianization_Hirata2003}), thereby improving the SNR of
        the estimated shear.
\end{itemize}

\section*{ACKNOWLEDGEMENTS}
\addcontentsline{toc}{section}{ACKNOWLEDGEMENTS}

XL thanks Erin Sheldon and Matthew Becker for their help on the LSST~DESC
simulations and discussions, Axel Guinot for the comments on the paper,
Masahiro Takada for discussions and hosting at IPMU, Richard Massey for
comments on the paper.
In addition, we thank the anonymous referee for feedback that improved the
quality of the paper.

XL and RM were supported by a grant from the Simons Foundation (Simons
Investigator in Astrophysics, Award ID 620789).

This paper makes use of software developed for the Vera C.\ Rubin Observatory.
We thank the Vera C.\ Rubin Observatory for making their code available as free
software at http://dm.lsst.org.

We thank the maintainers of \numpy{} \citep{numpy_Harris2020}, \texttt{SciPy}
\citep{scipy_Virtanen2020}, \texttt{Matplotlib} \citep{matplotlib_Hunter2007},
and \galsim{} \citep{GalSim} for their excellent open-source software.

\section*{DATA AVAILABILITY}
\addcontentsline{toc}{section}{DATA AVAILABILITY}

Our shear estimation project \texttt{AnaCal}, which is available from
\url{https://github.com/mr-superonion/AnaCal/} is targeted at developing
analytic tools for accurate weak lensing shear calibration which transform
measured galaxy properties such as ellipticity to shear distortion. To ensure
that our code development is more manageable, modular, and scalable, our code
is divided into two repositories. The code used to analyze simulated galaxy
images and obtain shape catalogs is available from
\url{https://github.com/mr-superonion/FPFS/}\,, and the code used for
calculating shear from shape catalogs is available from
\url{https://github.com/mr-superonion/impt/}\,.

\bibliographystyle{mnras}
\bibliography{citation}
\appendix

\section{LSST-like simulation}
\label{app:sim}

\renewcommand{\thefigure}{B\arabic{figure}}
\begin{figure}
\begin{center}
    \includegraphics[width=0.45\textwidth]{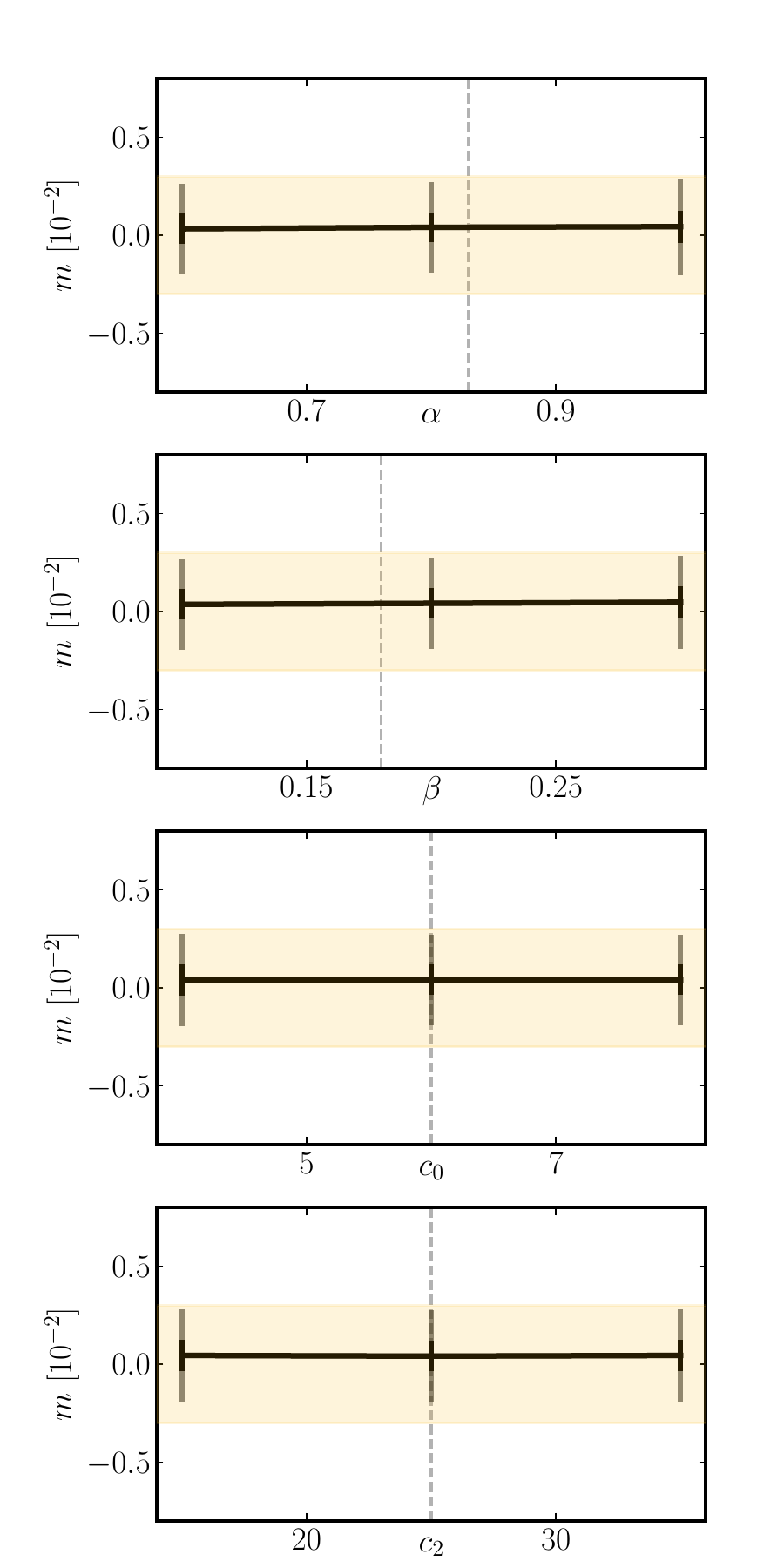}
\end{center}
\caption{
    The multiplicative bias as a function of hyperparameters ($\alpha$,
    $\beta$, $c_0$, $c_2$). We show the function in $1$D slices along these
    hyperparameters separately in four panels. The errorbars show the $1\sigma$
    and $3\sigma$ uncertainties, and the shaded region is the LSST ten-year
    requirement on the systematic control \citep{LSSTRequirement2018}. The
    vertical dashed lines indicate the optimal setups.
    }
    \label{fig:fpfs_ext_abc}
\end{figure}

We use the open-source package
\texttt{deswl-shear-sims}\footnote{\url{https://github.com/LSSTDESC/descwl-shear-sims}}
\citep{metaDet_LSST2023} to produce the LSST-like image simulations to test our
algorithm. The package uses
\texttt{WeakLensingDeblending}\footnote{\url{https://github.com/LSSTDESC/WeakLensingDeblending}}
package to generate our input galaxy sample, and galaxy images are rendered
with \galsim{} \citep{GalSim}. The galaxies' light profiles are approximated
with the best-fitting single S\'ersic model \citep{Sersic1963} or two-component
bulge-disk model \citep[with the bulge component
following][]{deVauProfile1948}.

We adopt LSST-like observational conditions: the image pixel scale is set to
$0\farcs2\,$, and the PSF image is modelled with a \cite{Moffat1969} profile,
\begin{equation}
    \label{Moffat PSF}
    p_{m}(\vx)=\left[1+c\left(\frac{|\vx|}
    {r_\mathrm{P}}\right)^2\right]^{-2.5},
\end{equation}
where $r_\mathrm{P}$ and $c$ are set such that the full width at half maximum
(FWHM) of the Moffat PSF is $0\farcs80$, matching the expected median seeing of
LSST images. The PSF model used in this paper is round; however, we note that
we tested the additive bias in \citet{FPFS_Li2023} using HSC-like PSF
\citep{HSC3_catalog} with ellipticity $(e_1=0.02, e_2 = -0.02)$ and find the
additive bias is less than $4\times 10^{-5}$. The noise level matches the LSST
ten-year observation. Specifically, we simulate images in the `$griz$' bands
with the same galaxy profiles and PSFs in each band without dithering;  while
the noise variance are different among them. Then we coadd the images in these
four bands with inverse variance weight of background noise. Since both the
galaxies and PSFs have the same profile in these bands and the weights are
constant over the images, the multi-band coadds have well-defined PSF
\citep{PSFCoadd_Mandelbaum2023}.

In order to tightly constrain shear biases with fewer simulations, we reduce
the shape noise from intrinsic galaxy shape and measurement error from image
noise in our tests for accuracy. Specifically, we generate two images for each
galaxy distorted by different shears --- ($g_1\!=\!0.02$,\, $g_2\!=\!0$) and
($g_1\!=\!-0.02$,\, $g_2\!=\!0$). However, the images share exactly the same
realisation of image noise \citep[following][]{preciseSim_Pujol2019,
metaDet_Sheldon2020}. In addition, we adopt the ring test setup
\citep{galsim_STEP2} by forcing our galaxy sample to contain galaxy pairs with
the same morphology and brightness but with the intrinsic major axes rotated by
$90 \deg$\,. Each simulated image covers 0.12 square degrees and contains about
100000 input galaxies. The input galaxy number density is about 230 per spare
arcmin. We measure multiplicative and additive bias in our shear estimator as
\begin{equation}
\label{eq:mbias_estimator}
    {m_1} = \frac{\langle \widehat{w e_1}^+ - \widehat{w e_1}^- \rangle}
    {0.02\left\langle \widehat{R_{1}}^{+} + \widehat{R_{1}}^{-} \right\rangle}
    -1\,,
\end{equation}
and
\begin{equation}
\label{eq:cbias_estimator}
    {c_1} = \frac{\langle \widehat{w e_1}^+ + \widehat{w e_1}^- \rangle}
    {\left\langle \widehat{R_{1}}^{+} + \widehat{R_{1}}^{-} \right\rangle}
\end{equation}
where $\widehat{w e_1}^+$ and $\widehat{R_{1}}^{+}$ are the first component of
the weighted ellipticity and its shear response after noise bias correction,
respectively. We refer readers to the sentences below
equation~\eqref{eq:shear_est_1} for their definitions. As indicated by the
superscript ``+'', they are estimated from the images distorted by the positive
shear, $(g_1\!=\!0.02,\, g_2\!=\!0)\,$. $\widehat{w e}_1^-$ and
$\widehat{R_{1}}^{-}$ are estimated from images distorted with the negative
shear, $(g_1\!=\!-0.02,\, g_2\!=\!0)$\,. Equations~\eqref{eq:cbias_estimator}
and \eqref{eq:mbias_estimator} assume that $\left\langle \widehat{R_{1}}^{+}
\right\rangle= \left\langle\widehat{R_{1}}^{-} \right\rangle$\, which is true
for our simulation since the input galaxy sample is the same for the images
with positively and negatively distorted galaxies. The galaxies in each
orthogonal galaxy pair and galaxies with different applied shears are selected
and weighted independently. After that, we apply our shear estimator to the
selected sample to test its performance. In addition, we divide our full galaxy
sample into $5000$ subfields (corresponding to about $590$ square deg in
total), and the errors on the means of multiplicative bias and additive bias
are estimated using the standard deviation of these $5000$ measurements.

\section{Stability of the estimator}
\label{app:stable}

We show the stability of our shear estimator with different hyperparameters
($\alpha$, $\beta$, $c_0$, $c_2$) in Fig.~\ref{fig:fpfs_ext_abc}. Specifically,
we check the multiplicative bias on the $1$D slices along these four
hyperparameters separately on a wide range covering the optimal parameter. The
results show that the accuracy is not significantly dependent on the
hyperparameters, and the multiplicative bias is below $3\times10^{-3}$ at the
$3\sigma$ level.

\label{lastpage}
\end{document}